\documentclass[conference]{IEEEtran}
\IEEEoverridecommandlockouts
\usepackage{cite}
\usepackage{comment}
\usepackage{amsmath,amssymb,amsfonts}
\usepackage{graphicx}
\usepackage{booktabs}
\usepackage{tabularx}
\usepackage{textcomp}
\usepackage{algorithm}
\usepackage{algpseudocode}
\usepackage{xcolor}
\usepackage{subcaption}
\usepackage[normalem]{ulem}
\usepackage{hyperref}
\def\BibTeX{{\rm B\kern-.05em{\sc i\kern-.025em b}\kern-.08em
    T\kern-.1667em\lower.7ex\hbox{E}\kern-.125emX}}
\definecolor{APcolor}{RGB}{255,0,0}
  
\begin{document}

\title{SCoPE: Subnetwork-Configurable Processing Elements for Continual Learning on Edge FPGAs%

\thanks{This work is co-funded by the ANR France 2030 PEPR-IA project, AdaptING Action (ANR-23-PEIA-0009), and by the European Union's Horizon Europe research and innovation program, COFUND SOUND.AI, under Marie Sk\l odowska-Curie Grant Agreement No.~101081674.}}

\author{
    \IEEEauthorblockN{Abdelghani Bourenane$^{1}$, Jean-Pierre David$^{2}$, Tarek Ould-Bachir$^{2}$, Julien
Denoulet$^{1}$,
Andrea Pinna$^{1}$}

    \IEEEauthorblockA{$^1$\textit{LIP6, Sorbonne Université, Paris, France}}
      \IEEEauthorblockA{$^2$\textit{Polytechnique Montréal, Montréal, Canada}}

    \IEEEauthorblockA{abdelghani.bourenane@sorbonne-universite.fr}
}

\maketitle

\begin{abstract}
In continual learning, neural networks must adapt their inference architecture dynamically to accommodate new tasks. However, traditional approaches, such as memory replay or regularisation-based methods adapt by updating network weights only, leaving both the neural architecture and the hardware data flow unchanged. In edge computing, where hardware resources are limited, this restricts both the total number of possible tasks and how quickly network parameters can be modified at an architectural level. We propose SCoPE (subnetwork-Configurable Processing Elements), an array of processing elements (PEs) that execute a unique sub-network for each task. Each PE can be configured at runtime using a 1-bit mask for each weight. This enables task-specific network connectivity while keeping the underlying 8-bit weights fixed. To minimize per-task mask storage, sub-network selection is restricted to the classification layers of a pretrained CNN. Experiments on Split-CIFAR-100 using an ImageNet-pretrained ResNet-18 achieve 89.8\% accuracy, matching full-precision end-to-end masking baselines while reducing weight size by a factor of 4 and mask storage overhead by a factor of 137. Implementing the proposed hardware architecture using High-Level Synthesis (HLS) on a PYNQ-ZU SoC FPGA achieves a 5.62$\times$ speedup in inference time compared to an optimised SIMD C implementation. Additionally, onboard integration with the Zynq System-on-Chip enables direct comparison with conventional full-weight updates.
In this case, our 1-bit-per-weight mask achieves a reduction in task adaptation latency of 7.61$\times$ compared to an 8-bit weight update.

\end{abstract}

\begin{IEEEkeywords}
Continual learning, dynamic architecture adaptation, Edge AI, Task-Adaptive Configurable Processing Element Array, SoC-FPGA
\end{IEEEkeywords}

\section{Introduction}

Biological brains exhibit lifelong learning, whereby acquiring one skill (e.g., playing the guitar) facilitates  the acquisition of another skill (e.g., playing the piano)~\cite{van2024continual}. In contrast, artificial neural networks suffer from catastrophic forgetting, whereby learning new tasks impairs performance on previously learned ones~\cite{kirkpatrick2017overcoming}. Continual Learning (CL) therefore seeks to address this limitation by enabling sequential learning from non-stationary data streams while balancing stability and plasticity under constrained computational resources~\cite{wang2024comprehensive}. CL scenarios are commonly categorized as domain-incremental, class-incremental, or task-incremental learning settings~\cite{van2019three}. To mitigate forgetting, existing CL approaches fall into three families~\cite{van2024continual}. 

\begin{itemize}
\item \textbf{Replay:} Interleave current-task data with stored samples from previous tasks (e.g., GEM~\cite{lopez2017gradient}, DER~\cite{buzzega2020dark}). 
\item \textbf{Regularization:} Penalize updates to critical parameters during fine-tuning
(e.g., EWC~\cite{kirkpatrick2017overcoming}, LwF~\cite{li2017learning}). 
\item \textbf{Architectural:} Allocate different tasks to disjoint subsets of parameters within a fixed network (e.g., PackNet~\cite{mallya2018packnet}) or a dynamically growing one (e.g., ProgNet~\cite{rusu2016progressive}). 
\end{itemize}

CL is fundamentally constrained by resource efficiency, necessitating customized hardware architectures that address both limited resources and the need for adaptability 
~\cite{kudithipudi2023design}. Such architectures must also address various design considerations (e.g., dataflow, external memory access, and computation), while offering reconfigurability. However, existing studies have primarily focused on replay or regularisation~\cite{park2025survey}, for which the neural
architecture remains fixed and the hardware dataflow is static
across tasks. Such approaches require the streaming of 
weight sets when adapting to new tasks~\cite{hayes2022online}.  
This leads to significant reconfiguration latency~\cite{bourenane2026reducing},   
and makes them unsuitable for safety-critical systems (e.g., autonomous vehicle perception~\cite{liu2019edge}). This constraint is further complicated by the fact that a single off-chip DRAM access consumes approximately $200\times$ the energy of a single Multiply-Accumulate (MAC) operation~\cite{chen2016eyeriss}, making architectures that minimize off-chip weight reloading during task reconfiguration essential for low-latency, energy-efficient continual adaptation at the edge~\cite{lesort2020continual}. Indeed, many FPGA deployments are limited to generic implementations that do not exploit architectural optimizations such as spatial parallelism, focusing solely on reducing replay memory size~\cite{aggarwal2023chameleon}. Additionally, some methods rely on regularization-based techniques, which introduce neither neural nor computational adjustments but instead focus solely on weight updates~\cite{karia2022scolar,elsayed2026streamlearn}. 
While architectural CL methods based on parameter allocation strategies \cite{mallya2018piggyback} are conceptually well aligned with the synaptic gating principles observed in biological systems~\cite{kudithipudi2022biological}, their deployment on hardware remains underexplored~\cite{piyasena2021accelerating,piyasena2020dynamically}. This creates a mismatch between current hardware implementations, which still treat task adaptation as a full-parameter update problem \cite{park2025survey}, and architectural CL methods, which aim to isolate task-specific sub-networks ~\cite{mallya2018piggyback}.

In this work, we propose a runtime-configurable hardware
architecture able to adapt to new tasks in a CL setting
by executing task-specific sub-networks. \\ Our contributions are as follows:

\begin{enumerate}

\item We propose a sub-network allocation method for the classifier, which operates on features extracted from a pre-trained CNN while using 8-bit quantized weights. Experiments on Split-CIFAR-100 using an ImageNet-pretrained ResNet-18 demonstrate 89.8\% accuracy, which is on par with full-precision, full-network masking, while requiring $137\times$ less mask size and four times less weight storage.

\item We present the classifier hardware architecture, SCoPE, a
configurable PE array that dynamically reconfigures the executed
sub-network at runtime through a 1-bit mask associated with each
weight. We implement the architecture using HLS and integrate it
into the programmable logic (PL) of the ZYNQ System-on-Chip.

\item We deploy SCoPE on a PYNQ-ZU MPSoC FPGA. Experiments show a $5.62\times$ improvement in inference time
compared with a SIMD C implementation on the embedded ARM
processor of the Processing System (PS). Our 1-bit-per-weight update achieves a $7.61\times$
reduction in task adaptation latency compared to an 8-bit weight
update.
\end{enumerate}
The remainder of this paper is organized as follows. Section~\ref{sec:relwork} reviews related work, Section~\ref{sec:sub-networks} presents sub-network-based CL, Section~\ref{sec:hardware} describes the proposed hardware architecture, and Section~\ref{sec:evaluation} reports the experimental results.

\section{Related Work}
\label{sec:relwork}
Replay- and regularization-based CL methods retain a common set
of model parameters across tasks, meaning updates for a new task
interfere with parameters important for previously learned ones ~\cite{wang2024comprehensive}.
Architectural parameter-allocation methods reduce this
interference by assigning different sub-networks to different tasks \cite{mallya2018piggyback}. SupSup~\cite{wortsman2020supermasks}
learns task-specific sub-networks within a fixed, randomly
initialized network, identifying each sub-network by learning
connection scores rather than modifying the underlying
weights~\cite{ramanujan2020s}. While these approaches demonstrate that
task adaptation can be achieved by selecting connections
rather than modifying weights, they are evaluated
algorithmically, without an equivalent hardware architecture.

Existing embedded CL implementations have targeted
replay-based methods,
retaining a fixed neural architecture and dataflow across
tasks~\cite{park2025survey}. LR~\cite{pellegrinilatent} freezes the early layers of a pretrained MobileNet backbone while only fine-tuning those after a chosen latent layer. Chameleon~\cite{aggarwal2023chameleon} builds on LR to optimize memory organization with a generic FPGA implementation. StreamLearn-FPGA~\cite{elsayed2026streamlearn} provides an MPSoC FPGA-based hardware
accelerator for EWC-based regularization \cite{kirkpatrick2017overcoming}.

Because these methods adapt by updating the full-precision
weights, task reconfiguration remains tied to weight streaming, 
and the hardware dataflow remains fixed across tasks, regardless
of hardware optimization. Table~\ref{tab:prior_hw_comparison}
summarizes these differences.
To our knowledge, per-task sub-network masking, which retains a fixed underlying parameter space while adapting only task-specific connectivity has not yet been explored as a dedicated hardware architecture. 
SCoPE closes this gap by reconfiguring task-specific connectivity directly in the PE array’s dataflow through a 1-bit mask per weight without streaming any weights.

\begin{table}[htbp]
\centering
\caption{Comparison with prior CL approaches. 
}
\label{tab:prior_hw_comparison}
\setlength{\tabcolsep}{3.5pt}
\small
\begin{tabular}{lccc}
\toprule
\textbf{Method} & \textbf{Adaptation} & \textbf{Reconfig.} & \textbf{Hardware} \\
                & \textbf{mechanism}  & \textbf{payload}   & \textbf{dataflow} \\
\midrule

StreamLearn~\cite{elsayed2026streamlearn} & Full network update& Full weight & Fixed \\
Chameleon~\cite{aggarwal2023chameleon} & Partial network update & Full weight & Fixed \\
SupSup~\cite{wortsman2020supermasks} & Masking, full network & 1 bit & -- \\
\textbf{SCoPE} & \textbf{Masking, classifier} & \textbf{1 bit} & \textbf{ Configurable} \\
\bottomrule
\end{tabular}
\end{table}

\section{sub-networks for Task-Incremental CL}
\label{sec:sub-networks}
\subsection{Sub-networks in a Randomly Initialized Network}
\label{subnet}
The theoretical foundation for finding sub-networks in fixed-weight, larger networks was introduced in prior work \cite{ramanujan2020s}. For a fully connected neural network (FNN) with layers
$\ell=1,\ldots,L$, layer $\ell$ contains $n_\ell$ neurons
$V^{(\ell)} = \{v_1,\ldots,v_{n_\ell}\}$. For a connection from
neuron $u \in V^{(\ell-1)}$ to neuron $v \in V^{(\ell)}$,
$w_{uv}$ is a fixed weight and $s_{uv}$ its learnable score, so
that the pre-activation of neuron $v$ is
$I_v = \sum_{u \in V^{(\ell-1)}} w_{uv} Z_u M_{uv}$, where $Z_u$
is the output of neuron $u$ and $M_{uv} \in \{0,1\}$ is the
connectivity mask. The weights are initialized once and kept
fixed: $w_{uv} \in \{-c,c\}$, $c = \sqrt{2/n_{\ell-1}}$.

For each task, $s_{uv}$ is associated with every connection.
During the forward pass, the scores are converted into a binary
mask by retaining the top $k\%$ of connections in each layer:
\begin{equation}
\label{scores}
M_{uv} = h(s_{uv}) =
\begin{cases}
1, & s_{uv} \text{ is in the top } k\% \text{ layer } \ell, \\
0, & \text{otherwise.}
\end{cases}
\end{equation}
The mask defines the task-specific sub-network while leaving the
underlying weights unchanged. Since the top-$k\%$ selection is
non-differentiable, a straight-through estimator (STE) is used
during backpropagation: the scores are updated while the fixed
weights are not, following
$s_{uv} \leftarrow s_{uv} - \alpha\frac{\partial\mathcal{L}}{\partial s_{uv}} \approx s_{uv} - \alpha\frac{\partial\mathcal{L}}{\partial I_v}w_{uv}Z_u$.
The layer-wise density $k_\ell$ follows a random graph formulation, in which each
connection is retained independently with probability:\begin{equation} k_\ell = \min\!\left(d_{\max},\ \varepsilon \cdot \frac{n_\ell + n_{\ell-1}}{n_\ell \, n_{\ell-1}}\right), \label{eq:er_sparsity_general} \end{equation} where $\varepsilon$ is a single global scale parameter shared across all layers, and $d_{\max}$ is a density cap. 

\subsection{Variant sub-network} 
\label{subsec:variant-sub-network} 
Edge devices have limited hardware resources~\cite{singh2024edge}, and supporting a large number of tasks on such devices requires moderating memory growth per task~\cite{park2025survey}. Since sub-network masks require dedicated storage per task and the size of the mask scales with the number of maskable parameters, we reduce the storage required per task by retaining the sub-network-selection mechanism of Section~\ref{subnet}, but restricting its application to the classifier rather than the entire network. Since the classifier accounts for only a small fraction of the network's total parameters, masking it alone yields a smaller per-task mask than masking the full network. Instead, the feature extractor is a fixed, pretrained network. This is motivated by the observation that representations
learned from large-scale visual datasets can be transferred effectively to downstream tasks, providing strong initialisation for new classification problems~\cite{yosinski2014transferable}. 
For the same hardware-efficiency reasons, the
classifier and feature extractor weights are quantized to 8-bit integers
(Section~\ref{subsec:quantization})

\begin{figure}
\centering
\includegraphics[width=\linewidth]{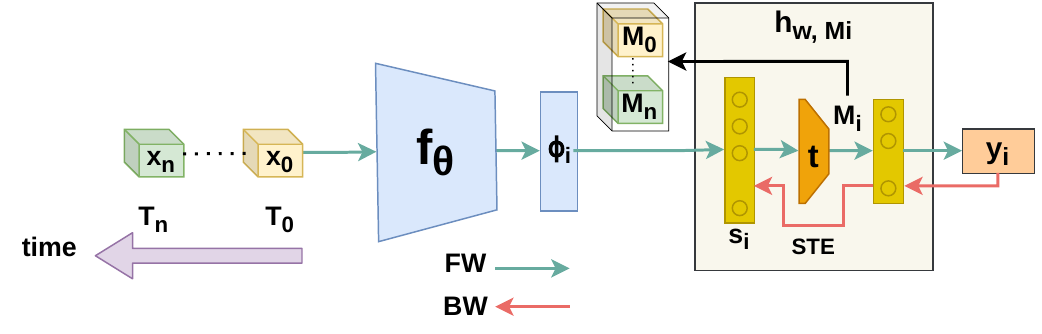}
\caption{Task-specific
sub-networks within a fixed-weight classifier operating on a frozen pretrained feature extractor.}
\label{fig:subnet-classifier}
\end{figure}

As illustrated in Fig.~\ref{fig:subnet-classifier}, the backbone is a fixed, pretrained feature extractor $f_\theta$, while sub-network selection is restricted to the classifier $h_{w,M_i}$. For tasks $T_0$ to $T_n$, a shared classifier weight set $w$ is retained, with each task selecting its own mask $M_0$ to $M_n$. Given input $x_i$, the feature vector $\phi_i = f_\theta(x_i)$ is passed to the classifier $h_{w,M_i}$, gated by the task-specific mask $M_i$, using the same scoring, thresholding, and STE update described in Section~\ref{subnet}.





\section{Proposed Hardware Architecture}
\label{sec:hardware}

\subsection{From Neural sub-networks to Task-Adaptive Configurable Processing Element Array}
\label{subsec:task-adaptive-PEs}\label{pe-arch}

\begin{figure}
\centering
\begin{subfigure}{0.48\textwidth}
        \includegraphics[width=\textwidth]{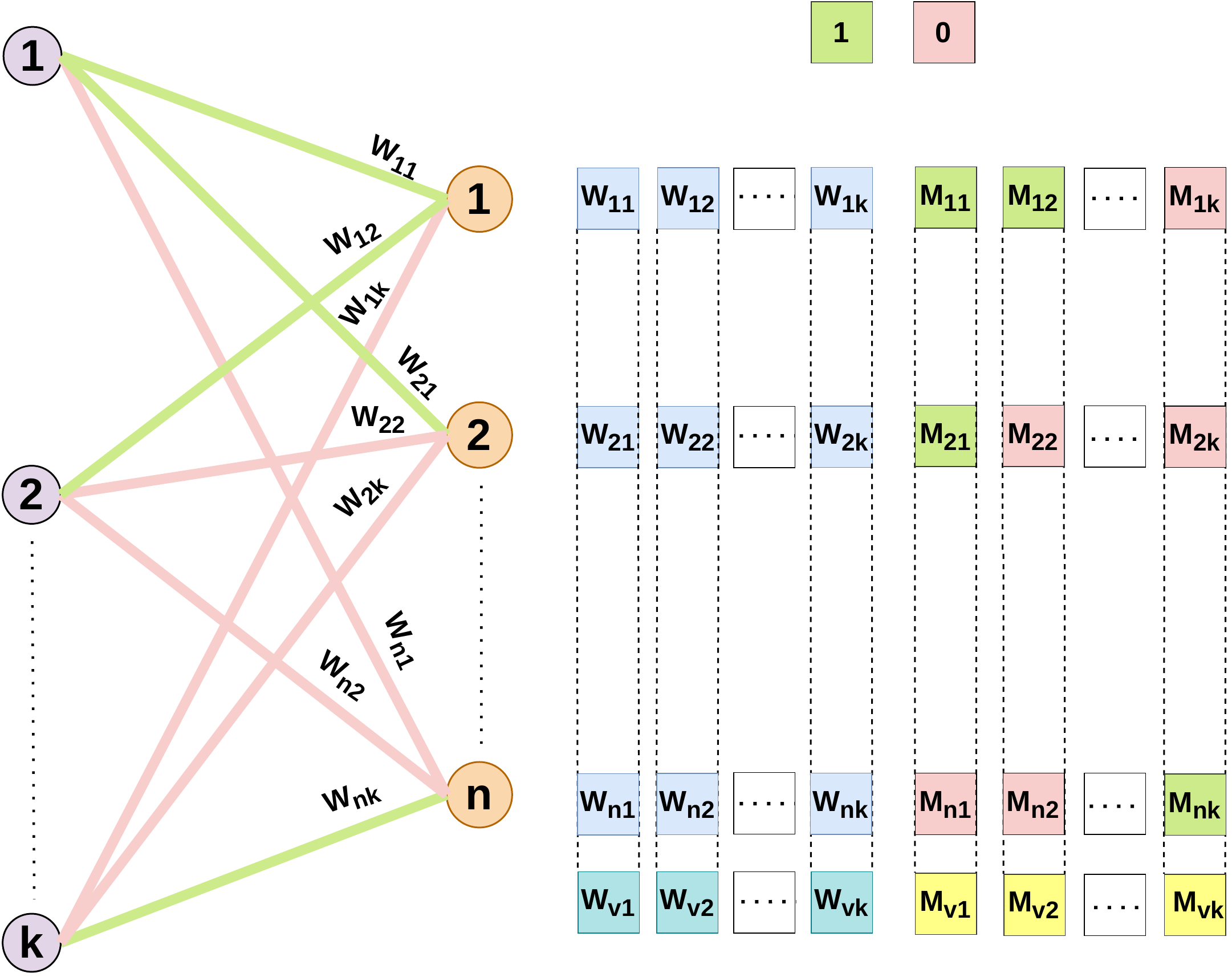}
    \caption{Neural configurations of sub-network of a task $T_i$}
    \label{fig:nn}
\end{subfigure}
\hfill

\begin{subfigure}{0.45\textwidth}
    \includegraphics[width=\textwidth]{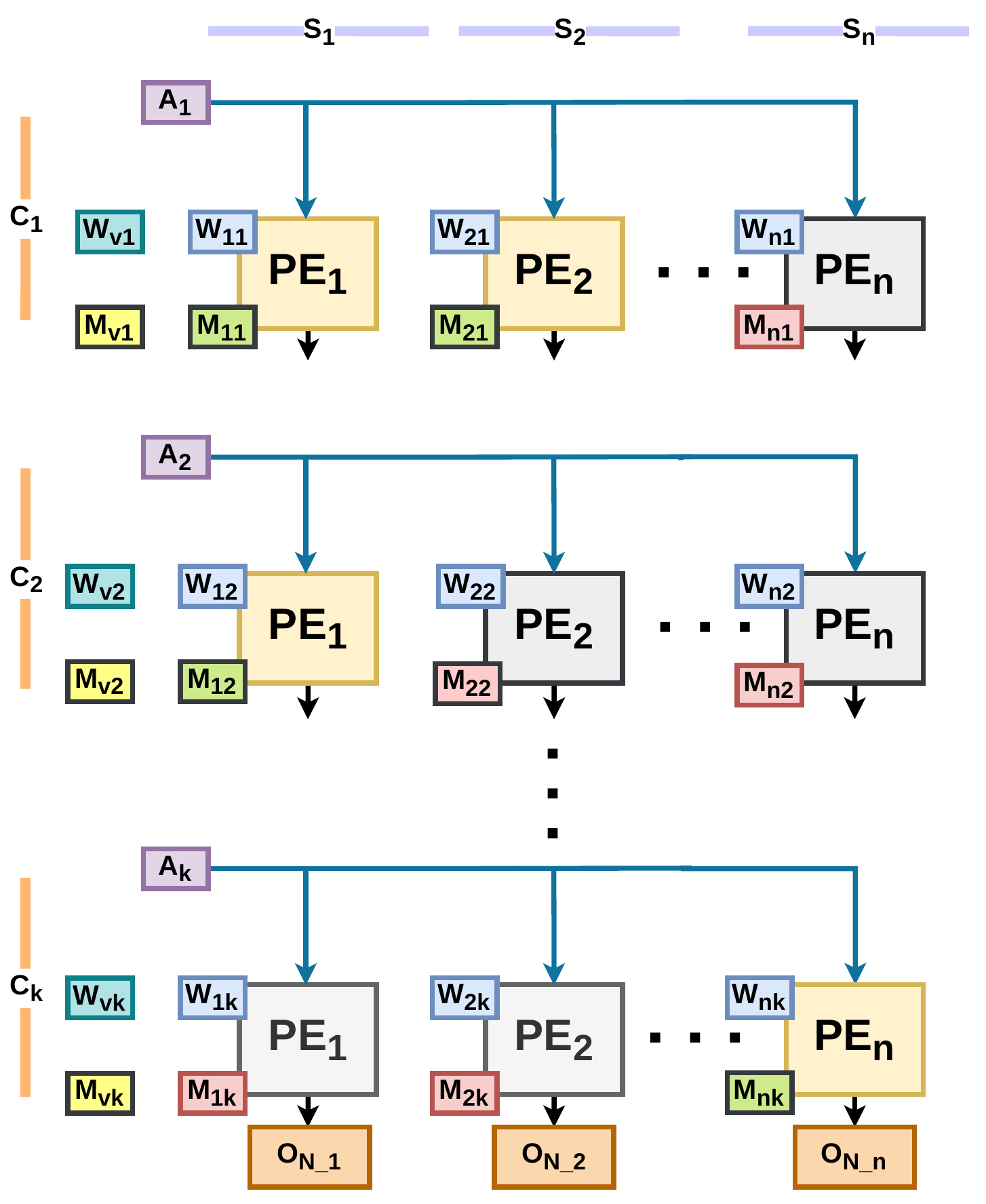}
    \caption{PE array runtime-configured for task $T_i$ subnetwork}
    \label{pe-array}
\end{subfigure}
        
\caption{Proposed hardware architecture}
\label{fig:pe-array}
\end{figure}

As illustrated in Fig.~\ref{fig:nn}, we organize the synaptic weight matrix $\mathbf{W} \in \mathbb{R}^{n \times k}$ along the network edges, connecting $k$ input neurons to $n$ output neurons. At a given task interval $t_i$, a task $T_i$ activates a specific sub-network defined by a binary masking matrix $\mathbf{M} \in \{0,1\}^{n \times k}$. The mask elements ($M_{11}$ to $M_{nk}$) follow the same indexing scheme as the weights ($W_{11}$ to $W_{nk}$), enabling a direct correspondence between weight storage and masking logic in hardware.
The weights are organized in vectors $W_{v1}$ to $W_{vk}$; this organization facilitates parallel computation and corresponds directly with the mask vectors $M_{v1}$ to $M_{vk}$, where $k$ indexes the input features.

To support this architecture, we design an array of PEs that exploits spatial parallelism. As illustrated in Fig.~\ref{fig:pe-array}, the proposed architecture consists of $n$ PEs, organized from $PE_{1}$ to $PE_{n}$. Each PE column consists of $n$ spatial elements ($S_1$ to $S_n$) that operate in parallel on a vectorized data stream, enabling SIMD-style (Single Instruction, Multiple Data) execution within the column. To minimize energy-intensive data movement, we employ a Weight-Stationary and Mask-Stationary data-flow. Weight and mask vectors $W_{vi}, M_{vi}$ are fetched, and their components are buffered locally within each PE register file, while an input feature $A_i$ is broadcast across the PE array, allowing multiple PEs to operate concurrently. Each PE is dynamically configurable via a task-specific binary mask, in which only the weights corresponding to active mask bits participate in the computation, while inactive weights are bypassed. Each PE computes the activation of an output neuron $O_{N_j}$ as a cumulative partial sum over $k$ cycles ($C_1$ to $C_k$), where each cycle represents the time to perform a MAC operation. For example, during $C_1$, vectors $W_{v1}, M_{v1}$ and their components $W_{11}$ to $W_{n1}$ and $M_{11}$ to $M_{n1}$ are distributed through the PE array. The computation is governed by the task-specific binary mask, with the core arithmetic operation run in the flow of Fig. \ref{pe-array} defined as:

\begin{equation}
O_{N_j} = \sum_{i=1}^{k} (W_{ij} \cdot X_i) \cdot M_{ij},
\label{eq:node_output}
\end{equation}

where $X_i$ denotes the $i$-th input feature, and $M_{ij}$ acts as a hardware-level gate. When $M_{ij} = 0$, the corresponding MAC operation is bypassed, preserving task-specific behavior without requiring on-chip weight reconfiguration.

\subsection{Quantization}
\label{subsec:quantization}

Weights are quantized after training using a signed 8-bit
fixed-point representation with per-layer symmetric,
max-magnitude scaling: for each layer $\ell$, a shift amount
$s_\ell$ is chosen such that $\text{round}(w \cdot 2^{s_\ell})$
fits within $[-127,127]$ for the layer's largest-magnitude
weight. All weights in that layer are scaled and rounded
accordingly. Masks remain unquantized by construction, as they
are already binary.
\subsection{HLS Architecture Design}
\label{subsec:hls-details}

Weights, masks, and accumulators are represented using a packed vector type, \texttt{DataPack<T,W>}~\cite{de2020transformations},
which stores $W$ elements of type $T$ contiguously within a
single wide \texttt{ap\_uint} register to enable single-transaction
memory access. Weights use $T{=}\texttt{int8\_t}$; masks use
$T{=}\texttt{ap\_uint<1>}$, 1 bit per element, with $W$
mask bits packed into a single $W$-bit-wide register. This packing is realized by \texttt{DataPack}'s underlying bit-range addressing.
(\texttt{data.range($(i+1) \cdot k_{\text{bits}} - 1$, $i \cdot k_{\text{bits}}$)}), which places element $i$ at bit position $i$ of the register for $k_{\text{bits}} = 1$.
On the host side, mask arrays are correspondingly packed 8 elements per byte prior to transfer,
matching this in-register layout. Partial sums accumulate in
$T{=}\texttt{int16\_t}$ to avoid overflow across the reduction in
Eq.~\eqref{eq:node_output}.

The vector width $W$ is fixed for a given deployment. For
layers whose output count $n$ is smaller than $W$, only the
first $n$ positions are computed. In storage, the corresponding
weight and mask arrays are explicitly zero-padded to the full
width $W$; in compute, the remaining $W{-}n$ output positions
are not written by the unrolled reduction. 

Realizing Eq.~\eqref{eq:node_output}, each of a
layer's $n$ output neurons computes independently and in
parallel. The reduction over the $k$ input features is staged
through a fully-partitioned local accumulator array. 
An \texttt{HLS UNROLL} directive instantiates $n$ parallel PEs
(Section~\ref{pe-arch}) over this array, with one PE per output lane. 
Each PE realizes a
masked MAC unit and the result is packed into the vector
representation once per reduction step. Each PE realizes the mask gate as a multiplexer select, where the
mask bit  selects either the weight value or a hardwired zero
($w_{ij}^{\text{gated}} = M_{ij} \;?\; w_{ij} : 0$). An inactive
connection contributes a zero operand to the
accumulation, so the set of connections carrying signal through
the array is determined by the mask loaded at runtime.
Thus, the same physical array realizes a different dataflow graph
for each task, without any resynthesis or modification to the
stored weights. The outer reduction over the $k$ input features,
is
pipelined with an initiation interval of one cycle
(\texttt{HLS PIPELINE II=1}), allowing a new input feature to be
consumed every clock cycle. Input features $X_i$ are
staged into a local input buffer via a burst-inferred
\texttt{memcpy} prior to the reduction, ensuring a single wide
AXI transaction rather than $k$ individual reads.

Following accumulation, each layer applies a fixed-point rescale
($y \gg s_\ell$, Section~\ref{subsec:quantization}) and, for
intermediate layers, a ReLU nonlinearity. The final layer's raw
outputs are passed directly to an argmax operation for
classification. Each layer's $n$-wide PE array is synthesized as a separate, spatially-instantiated hardware block.
\subsection{Integration into SoC FPGA}\label{systemcl}

\begin{figure}
    \centering
    \includegraphics[width=\linewidth]{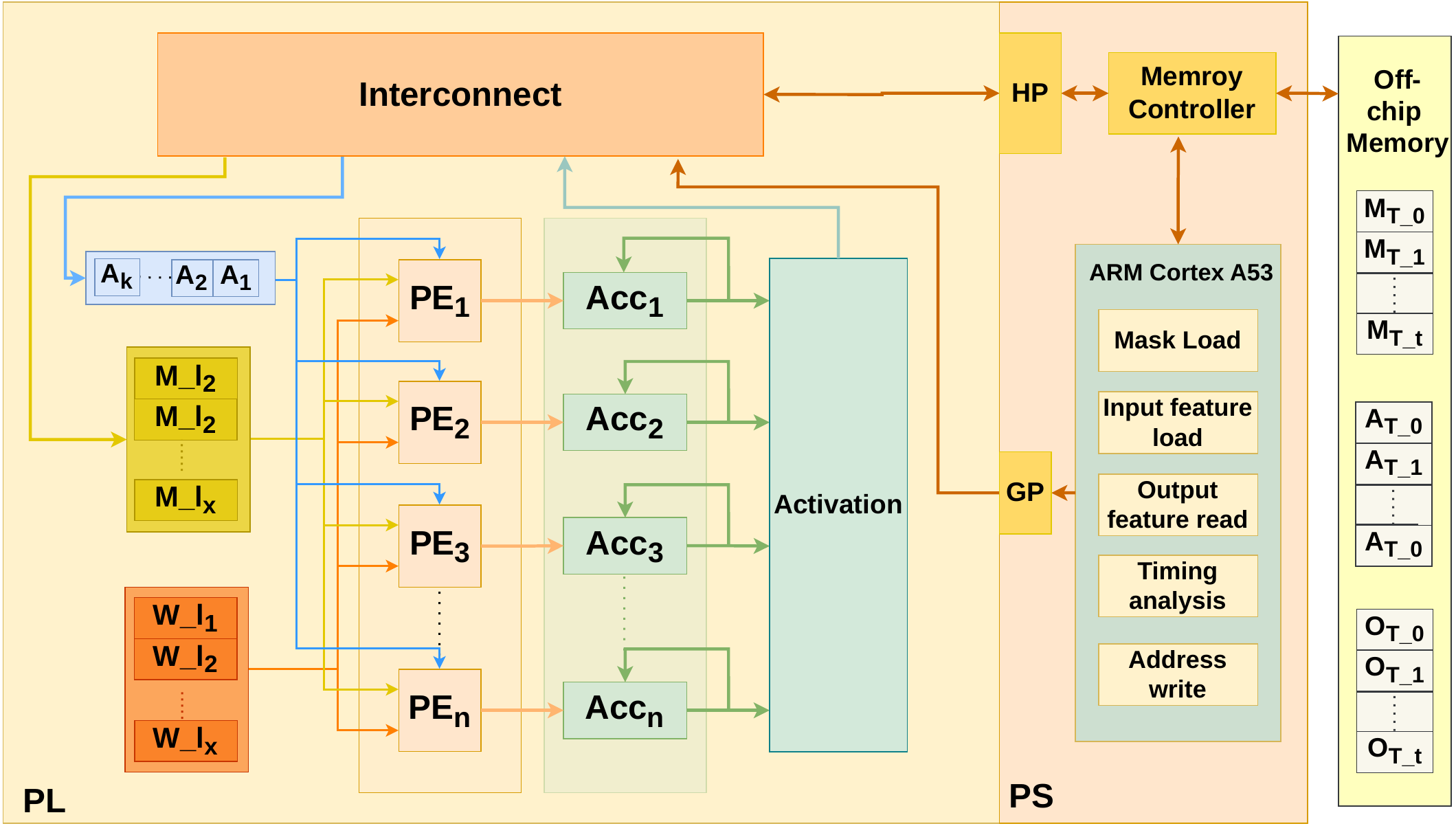}
    \caption{MPSoC-FPGA-based CL adaptive system }
    \label{fig:psoc}
\end{figure}

The HLS kernel has been implemented and integrated into a ZYNQ SoC FPGA platform.  
As shown in Fig.~\ref{fig:psoc}, the PE array is implemented in the PL, where it performs masked weight–input multiplications.  
Associated accumulators then store the intermediate partial sums for each output neuron. Each PE receives an input feature $A_{_i}$ from an input buffer. Task-specific masks are buffered per network layer ($M_{l1}$ to $M_{lx}$), while the corresponding weights are stored in on-chip BRAM as $W_{l1}$ to $W_{lx}$. Each PE ($PE_1$ to $PE_n$) writes its result to a dedicated accumulator ($Acc_1$ to $Acc_n$). The PE engine then produces a vector of output activations, which are forwarded to the activation module. 

We have created AXI memory-mapped interfaces for the PE kernel, which allow access to off-chip DDR memory. These master interfaces handle input features ($A$), task masks ($M$), and output features ($O$), and are all connected to a central interconnect. The interconnect synchronizes all master interfaces accessing off-chip memory via the slave High-Performance (HP) port of 64-bits width. The HP port provides a read channel for fetching input features; and a write data channel for storing output results back to memory. To enable efficient burst transfers, binary mask values are packed into 64-bit words.  Off-chip memory stores a bank of task-specific masks, with the PS specifying the base address for each task. The ARM Cortex-A53 CPU, in particular, uses the general-purpose (GP) port to configure the base addresses of the kernel master interfaces and to control kernel execution. The GP port connects to the interconnect which provides the associated control channel used to transmit configuration and control data. The CPU writes task masks ($M_{T_1}$ to $M_{T_t}$) to off-chip memory, loads the corresponding input features ($A_{T_1}$ to $A_{T_t}$) and retrieves the output features ($O_{T_1}$ to $O_{T_t}$) for evaluation.

\section{Evaluation}
\label{sec:evaluation}

\subsection{Experimental methodology}
\subsubsection{Dataset} Following relevant prior work \cite{kang2024continual}, we benchmark SCoPE on Split-CIFAR-100 dataset under a task-incremental setup. This dataset includes 60,000 $32 \times 32$ RGB images spanning 100 object classes. We partition the 100 classes into 20 sequential groups, with each task introducing 5 distinct classes. During training, tasks are presented sequentially, and the model has no access to raw historical data from previously encountered tasks. 

\subsubsection{Model}
We employ the ResNet18 architecture as the feature extractor $f_\theta$, using its publicly released
ImageNet-pretrained weights. The classification layer of ResNet18
is discarded and the 512-dimensional feature vector produced by the
global average pooling layer is used as $\phi_i$, the
input to the task-specific masked classifier $h_{w,M_i}$, which
has two hidden layers of 128 neurons each and an output layer of
5 neurons. 

\subsection{Accuracy}
Experiments were run on an NVIDIA RTX 3090 (24\,GB) GPU. We evaluate the cumulative average accuracy after training
through task $k$, defined as:
\begin{equation}
\text{Acc}_k = \frac{1}{k} \sum_{i=1}^{k} A_{k,i} \label{eq:cumavg}
\end{equation}

where $A_{k,i}$ denotes accuracy on task $i$ using the model
obtained after training through task $k$. As a baseline, we implement sequential fine-tuning: a
classifier, operating on features from the same pretrained, feature extractor used by SCoPE, is trained continuously
across all tasks in sequence. 

\begin{figure}
    \centering
    \includegraphics[width=0.8\linewidth]{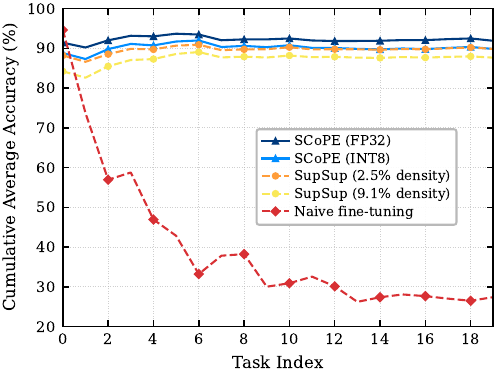}
    \caption{Cumulative average accuracy across, encountered tasks}
    \label{fig:per_task_all_conditions}
\end{figure}

Fig.~\ref{fig:per_task_all_conditions} reports cumulative average
accuracy for SCoPE with the pretrained-backbone variant used here
(FP32 and INT8) with classifier density $50\%$, SupSup~\cite{wortsman2020supermasks} at two mask
densities set via Eq.~\eqref{eq:er_sparsity_general}, and the sequential fine-tuning baseline. All
masking-based configurations retain a dedicated, frozen mask per
task and stabilize within 5--7 tasks with zero forgetting by
construction; fine-tuning, lacking any such isolation mechanism,
collapses from 94.6\% to 27.4\% over the same period,
demonstrating the catastrophic forgetting 
that
masking prevents. The pretrained-backbone configuration SCoPE uses outperforms SupSup's best result (89.9\% at 2.5\% density) by 2.0 points, 
consistent with pretrained representations transferring more effectively than random-backbone features~\cite{kornblith2019better}. 
SupSup's accuracy falls by 2.2 points as density rises from 2.5\% to 9.1\%. 
Notably, SupSup achieves a level of accuracy comparable to that of SCoPE while activating only 2.5\% of its randomly initialised network per task. This demonstrates that extreme sparsity alone can approach the accuracy of a fully dense, pre-trained backbone.
However, as SupSup masks the entire network and SCoPE only the classifier, their per-task storage costs differ substantially. SupSup's mask spans all 11.28M network parameters, whereas SCoPE's spans only the 82{,}560-parameters classifier, a reduction in per-task mask storage by a factor of 137.

Quantizing to INT8 (backbone via PyTorch's standard post-training quantization and classifier via Section~\ref{subsec:quantization}) costs 2.1 points (91.9\%$\to$89.8\%)
. Nevertheless, the quantized result essentially matches SupSup's best full-precision configuration, indicating that the benefit of pretraining's accuracy largely offsets the cost of INT8 required for the hardware implementation in Section~\ref{sec:hardware}.

\subsection{Hardware Evaluation}
\subsubsection{Hardware Resource and Performance}
\label{sec:execution-time}

The HLS design presented in Section~\ref{subsec:hls-details} was
implemented using AMD Vitis HLS 2024.2. Table~\ref{tab:layer_resources}
reports the post-synthesis resources utilization for the proposed
classifier on a Zynq UltraScale+ XCZU5EV device (PYNQ-ZU board),
targeting a 100~MHz clock frequency. The design reflects
the microarchitecture of Section~\ref{subsec:task-adaptive-PEs}: each
layer instantiates \(n\) parallel PEs, one per output neuron, and
each PE is mapped to a dedicated DSP block. Consequently, the usage of DSP scales directly with the number of output neurons per layer. For Layers 1 through 3, these numbers are 128, 128 and 5, respectively,
confirming the spatial parallelism described in
Section~\ref{subsec:hls-details}.

The per-layer compute latency follows directly from this
structure. Each PE accumulates its output over \(k\) cycles, one
masked MAC per input feature, as defined in
Eq.~\eqref{eq:node_output}, so the reduction depth equals the
number of input features \(k\). With the outer reduction pipelined
at II=1, the measured latency is \(k + 5\) cycles, where the
5-cycle constant corresponds to pipeline fill and drain. This confirms that the
HLS implementation preserves the intended compute schedule, with
latency scaling linearly in the input-feature dimension and
independent of the output-neuron count, which is processed
entirely in parallel.


\begin{table}[htbp]
\centering
\caption{PEs compute latency and resource usage per layer}
\label{tab:layer_resources}
\begin{tabularx}{\linewidth}{Xccccc}
\toprule
\textbf{L} & \textbf{Latency} & \textbf{BRAM} & \textbf{DSP} & \textbf{LUT} & \textbf{FF} \\
\midrule
  1 & 517 & 33 (11\%) & 128 (10\%) & 4{,}333 (3\%) & 4{,}322 (1\%) \\
 2 & 133 & 33 (11\%) & 128 (10\%) & 4{,}327 (3\%) & 4{,}318 (1\%) \\
 3 & 133 & 2 (1\%)   & 5 (1\%)    & 529 (1\%)     & 4{,}656 (1\%) \\
\midrule
\textbf{Total} & \textbf{783} & \textbf{73 (25\%)} & \textbf{261 (20\%)} & \textbf{21{,}025 (17\%)} & \textbf{19{,}249 (8\%)} \\
\bottomrule
\end{tabularx}
\end{table}

Power evaluation using Vivado Power Analyzer post-synthesis
reports a total dynamic power of 0.323~W for the classifier IP
At the 783-cycle 
compute
schedule established above, this corresponds to an active compute
energy of 2.53~$\mu$J per feature vector.
\subsubsection{Execution Time}

To benchmark hardware acceleration performance, software baselines were implemented in Python and C on the ARM Cortex-A53 CPU. Against these baselines, FPGA speedup reaches $41.2\times$ (scalar C) and $82.6\times$ (Python/NumPy); however, these figures are inflated by non-vectorized-loop overhead rather than reflecting a fair comparison.

\begin{figure}[t]
    \centering
    \includegraphics[width=0.85\linewidth]{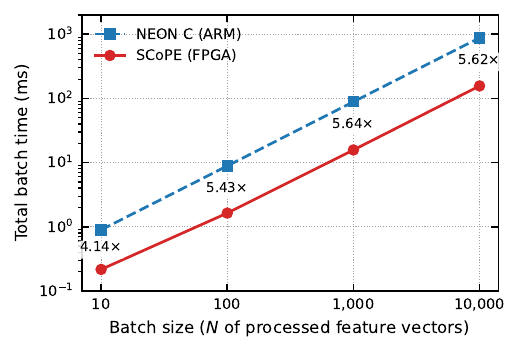}
    \caption{Total batch execution time and FPGA speedup}
    \label{fig:exec-time}
\end{figure}

Therefore, we developed an optimized 128-bit
SIMD NEON C implementation of the classifier, running 16 8-bit
MACs in parallel~\cite{martinotlagarde2025how}, which we executed on the ARM Cortex-A53 of the PYNQ-ZU board.
Fig.~\ref{fig:exec-time} reports total batch execution time and
FPGA speedup across four batch sizes ($N=10$ to $10{,}000$
feature vectors). At $N=10$, fixed AXI overhead is a noticeable fraction of runtime, limiting speedup to
$4.14\times$ over NEON C. Beyond $N=100$, compute time dominates
this overhead, and per-vector latency stabilizes near
$15$--$16\,\mu\text{s}$ for the FPGA and $88\,\mu\text{s}$ for
NEON C, yielding a stable $5.4\times$--$5.6\times$ speedup,
reaching $5.62\times$ at $N=10{,}000$.

\subsubsection{Configuration latency}
\begin{figure}
    \centering
    \includegraphics[width=0.8\linewidth]{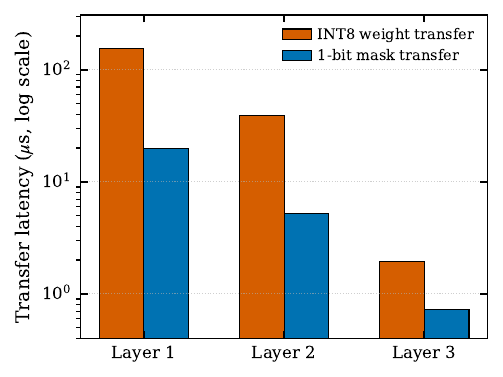}
    \caption{Network configuration latency}
    \label{fig:reconfig_latency}
\end{figure}

Existing CL hardware approaches that adapt a network to a new
task require writing an updated
set of weights to the deployed model for each new task \cite{li2025unleashing}. In
contrast, our proposed approach never writes to the shared weight
tensor $w$: $w$ remains fixed after initialization, and task
adaptation is achieved through the mask $M_i$.
Consequently, task reconfiguration in our system only requires the mask to be transferred, with no weight write-back. We evaluate the transfer latency of the proposed 1-bit mask against state-of-the-art MPSoC-FPGA CL architectures that require full 8-bit weight updates, such as Chameleon~\cite{aggarwal2023chameleon} and StreamLearn-FPGA~\cite{elsayed2026streamlearn}, using an identical 64-bit AXI port configuration across both schemes. The total
transfer latency, $T_{\text{total}}$, is modeled based on our
Logic Analyzer experiments of AXI handshaking behavior as follows:
\begin{equation}
T_{\text{total}} = L_{\text{init}} + (N_{\text{burst}} - 1) \times (B_{\text{size}} + G) + B_{\text{size}}
\label{eq:latency}
\end{equation}
where $L_{\text{init}}=57$ cycles represents the address-to-data
latency 
, $G$ is the
14-cycle inter-burst delay, and $N_{\text{burst}}$ is the number
of required AXI bursts, with burst size $B_{\text{size}}=16$.
Fig.~\ref{fig:reconfig_latency} reports the resulting latency for
each classifier layer. The achievable speedup tracks the
$8\times$ reduction in transferred data volume most closely for
larger layers, where the payload dominates $T_{\text{total}}$
(Layer~1: $7.85\times$), while fixed per-burst overhead erodes
the benefit for the smallest layer (Layer~3: $2.64\times$).
Summed across all three layers, total reconfiguration latency is
$25.59~\mu s$ for the proposed 1-bit mask versus $194.79~\mu s$
for INT8 weights. Accounting for the 0.069~W AXI interconnect
power estimated by Vivado Power Analyzer for this data path, the
corresponding transfer energy is $1.77~\mu$J for the mask versus
$13.44~\mu$J for INT8 weights. Both metrics yield the same
$7.61\times$ reduction, as expected under the assumption of
constant interconnect power.
\section{Conclusion}

In this work, we introduced SCoPE (subnetwork-Configurable Processing Elements), a novel hardware architecture featuring a runtime-configurable dataflow tailored for continual learning (CL) scenarios. By constraining sub-network allocation to the classifier layers and leveraging 8-bit quantized weights, SCoPE achieves \textbf{89.8\%} accuracy on the Split-CIFAR-100 benchmark, matching the performance of full-precision, full-network masking baselines. Notably, this is accomplished with a \textbf{137 $\times$ reduction in mask storage} and a \textbf{4 $\times$ reduction in weight storage}.

The core of SCoPE lies in its configurable Processing Element (PE) array, which dynamically applies weight and mask vectors to execute task-specific sub-networks. Through the use of 1-bit-per-weight masks, our approach \textbf{reduces the latency for updating task configurations by 7.61 $\times$} compared to traditional 8-bit weight updates. Furthermore, by exploiting hardware parallelism via FPGA-based architecture, SCoPE delivers a \textbf{5.62$\times$ speedup} in inference time over an optimized SIMD implementation on an embedded ARM processor.

These results underscore the potential of SCoPE as an efficient and scalable solution for CL in resource-constrained edge environments. Future work may explore extending this architecture to deeper networks or additional layers, further broadening its applicability to a wider range of CL tasks.

\bibliographystyle{IEEEtran}
\bibliography{bib}
\end{document}